\documentclass{aa}

\usepackage{graphicx}

\usepackage[varg]{txfonts}
\usepackage{natbib}
\bibpunct{(}{)}{;}{a}{}{,}
\usepackage{array}
\usepackage{xspace}
\usepackage{lscape}
\usepackage{url}
\usepackage{arydshln}
\usepackage[hyperfootnotes=false, linktocpage=true, breaklinks=true, colorlinks=true, linkcolor=blue, citecolor=blue, urlcolor=blue]{hyperref}
\usepackage[all]{hypcap}
\usepackage{longtable}
\usepackage{afterpage}

\usepackage[dvipsnames]{xcolor}

\newcommand{\fcov}{{\ensuremath{C_f}}\xspace}
\newcommand{\FeKa}{Fe K\ensuremath{\alpha}\xspace}

\newcommand{\kms}{\ensuremath{\mathrm{km\ s^{-1}}}\xspace}
\newcommand{\NH}{\ensuremath{N_{\mathrm{H}}}\xspace}

\newcommand{\xmm}{{XMM-\it{Newton}}\xspace}
\newcommand{\chandra}{{\it Chandra}\xspace}
\newcommand{\swift}{{\it Swift}\xspace}

\newcommand{\hst}{{HST}\xspace}

\newcommand{\ngc}{{NGC~3227}\xspace}

\newcommand{\nustar}{{\it NuSTAR}\xspace}

\newcommand{\ergflux}{{\ensuremath{\rm{erg\ cm}^{-2}\ \rm{s}^{-1}}}\xspace}
\newcommand{\ergs}{{\ensuremath{\rm{erg\ s}^{-1}}}\xspace}
\newcommand{\cm}{{\ensuremath{\rm{cm}^{-2}}}\xspace}

\newcommand{\cloudy}{{\tt Cloudy}\xspace}
\newcommand{\spex}{\xspace{\tt SPEX}\xspace}
\newcommand{\pion}{\xspace{\tt pion}\xspace}

\newcommand{\ebv}{\ensuremath{{E(B-V)}}\xspace}

\mathchardef\mhyphen="2D

\begin{document}

\title{Transient obscuration event captured in NGC 3227}
\subtitle{I. Continuum model for the broadband spectral energy distribution}

\author{
M. Mehdipour \inst{1,2}
\and
G.A. Kriss \inst{1}
\and
J.S. Kaastra \inst{2,3}
\and
Y. Wang \inst{2,3,4,5}
\and
J. Mao \inst{6,2}
\and
E. Costantini \inst{2,7}
\and
N. Arav \inst{8}
\and
E. Behar \inst{9}
\and
\newline
S. Bianchi \inst{10}
\and
G. Branduardi-Raymont \inst{11}
\and
M. Brotherton \inst{12}
\and
M. Cappi \inst{13}
\and
B. De Marco \inst{14}
\and
L. Di Gesu \inst{15}
\and
J. Ebrero \inst{16}
\and
\newline
S. Grafton-Waters \inst{11}
\and
S. Kaspi \inst{17}
\and
G. Matt \inst{10}
\and
S. Paltani \inst{18}
\and 
P.-O. Petrucci \inst{19}
\and
C. Pinto \inst{20}
\and
G. Ponti \inst{21,22}
\and
F. Ursini \inst{13}
\and
\newline
D.J. Walton \inst{23}
}
\institute{
Space Telescope Science Institute, 3700 San Martin Drive, Baltimore, MD 21218, USA \\ \email{mmehdipour@stsci.edu}
\and
SRON Netherlands Institute for Space Research, Sorbonnelaan 2, 3584 CA Utrecht, the Netherlands
\and
Leiden Observatory, Leiden University, PO Box 9513, 2300 RA Leiden, the Netherlands
\and
CAS Key Laboratory for Research in Galaxies and Cosmology, Department of Astronomy, University of Science and Technology
of China, Hefei 230026, China
\and
School of Astronomy and Space Science, University of Science and Technology of China, Hefei 230026, China
\and
Department of Physics, University of Strathclyde, Glasgow G4 0NG, UK
\and
Anton Pannekoek Institute, University of Amsterdam, Postbus 94249, 1090 GE Amsterdam, The Netherlands
\and
Department of Physics, Virginia Tech, Blacksburg, VA 24061, USA
\and
Department of Physics, Technion-Israel Institute of Technology, 32000 Haifa, Israel
\and
Dipartimento di Matematica e Fisica, Universit\`{a} degli Studi Roma Tre, via della Vasca Navale 84, 00146 Roma, Italy
\and
Mullard Space Science Laboratory, University College London, Holmbury St. Mary, Dorking, Surrey, RH5 6NT, UK
\and
Department of Physics and Astronomy, University of Wyoming, Laramie, WY 82071, USA
\and
INAF-IASF Bologna, Via Gobetti 101, I-40129 Bologna, Italy
\and
Departament de Física, EEBE, Universitat Politècnica de Catalunya, Av. Eduard Maristany 16, E-08019 Barcelona, Spain
\and
Italian Space Agency (ASI), Via del Politecnico snc, 00133, Roma, Italy
\and
Telespazio Vega UK for the European Space Agency (ESA), European Space Astronomy Centre (ESAC), Camino Bajo del Castillo,s/n, E-28692 Villanueva de la Ca\~nada, Madrid, Spain
\and
School of Physics and Astronomy and Wise Observatory, Tel Aviv University, Tel Aviv 69978, Israel
\and
Department of Astronomy, University of Geneva, 16 Ch. d'Ecogia, 1290 Versoix, Switzerland
\and
Univ. Grenoble Alpes, CNRS, IPAG, 38000 Grenoble, France
\and
INAF-IASF Palermo, Via U. La Malfa 153, I-90146 Palermo, Italy
\and
INAF-Osservatorio Astronomico di Brera, Via E. Bianchi 46, I-23807 Merate (LC), Italy
\and
Max Planck Institute fur Extraterrestriche Physik, 85748, Garching, Germany
\and
Institute of Astronomy, University of Cambridge, Madingley Road, Cambridge CB3 0HA, UK
}
\date{Received 17 May 2021 / Accepted 24 June 2021}
\abstract {
From \swift monitoring of a sample of active galactic nuclei (AGN) we found a transient X-ray obscuration event in Seyfert-1 galaxy NGC 3227, and thus triggered our joint \xmm, \nustar, and Hubble Space Telescope (\hst) observations to study this event. Here in the first paper of our series we present the broadband continuum modelling of the spectral energy distribution (SED) for NGC 3227, extending from near infrared (NIR) to hard X-rays. We use our new spectra taken with \xmm, \nustar, and the HST Cosmic Origins Spectrograph (COS) in 2019, together with archival unobscured \xmm, \nustar, and HST Space Telescope Imaging Spectrograph (STIS) data, in order to disentangle various spectral components of \ngc and recover the underlying continuum. We find the observed NIR-optical-UV continuum is explained well by an accretion disk blackbody component (${T_{\rm max} = 10}$~eV), which is internally reddened by ${E(B - V) = 0.45}$ with a Small Magellanic Cloud (SMC) extinction law. We derive the inner radius ($12\, R_{{\rm{g}}}$) and the accretion rate (0.1~$M_{\odot}$ yr$^{-1}$) of the disk by modelling the thermal disk emission. The internal reddening in \ngc is most likely associated with outflows from the dusty AGN torus. In addition, an unreddened continuum component is also evident, which likely arises from scattered radiation, associated with the extended narrow-line region (NLR) of \ngc. The extreme ultraviolet (EUV) continuum, and the `soft X-ray excess', can be explained with a `warm Comptonisation' component. The hard X-rays are consistent with a power-law and a neutral reflection component. The intrinsic bolometric luminosity of the AGN in \ngc is about ${2.2 \times 10^{43}}$~\ergs in 2019, corresponding to 3\% Eddington luminosity. Our continuum modelling of the new triggered data of \ngc requires the presence of a new obscuring gas with column density ${\NH = 5 \times 10^{22}}$~\cm, partially covering the X-ray source ($\fcov = 0.6$).
}
\authorrunning{M. Mehdipour et al.}
\titlerunning{Transient obscuration event captured in NGC 3227. I.}
\keywords{
X-rays: galaxies -- galaxies: active -- galaxies: Seyfert -- galaxies: individual: NGC 3227 -- accretion disks -- techniques: spectroscopic 
\vspace{-0.5cm}}
\maketitle
\section{Introduction}
\label{intro_sect}
Long-term monitoring of X-ray spectral variability in type-1 AGN with the Rossi X-ray Timing Explorer (RXTE) and \swift observatories have revealed that sometimes they display sudden spectral hardening events (see e.g. \citealt{Mark14,Kaas14,Mehd17}). These events are seen as spikes in the X-ray hardness ratio lightcurves, which typically last for days or weeks, or even several years in the case of \object{NGC~5548}. Over the past few years new results regarding the nature and origin of these transient spectral hardening events in type-1 AGN suggest they are caused by obscuring winds from the accretion disk, as first discovered in NGC~5548 \citep{Kaas14}. Simultaneous multi-wavelength spectroscopy in the UV and X-ray energy bands has been instrumental for the study of these events. Transient X-ray obscuration in AGN is found to appear with an associated UV broad absorption-line component \citep{Kris19}, and in some cases with a high-ionisation component in the Fe K-band \citep{Mehd17}.

In 2015 we started a monitoring program with the {\it Neil Gehrels Swift Observatory} \citep{Gehr04} to observe the X-ray hardness variability in a sample of Seyfert-1 AGN, in order to trigger joint ToO observations with \xmm \citep{Jans01}, \nustar \citep{Harr13}, and the Hubble Space Telescope (HST) Cosmic Origins Spectrograph (COS, \citealt{Green12}). In December 2016 we found an event in NGC~3783 and triggered our observations. \citet{Mehd17} show that the spectral hardening is caused by an obscuring wind, outflowing with a velocity of few thousand \kms, that crosses our line of sight. These obscuring winds are remarkably different from the commonly-seen warm-absorber outflows in AGN (e.g. \citealt{Blu05,Laha14}). They have higher column densities and outflow velocities, and are located in the broad-line region (BLR) rather the narrow-line region (NLR). The optically-thick obscuration shields much of the X-ray radiation, which has important impacts on the ionisation state and our interpretation of both the warm absorber \citep{Arav15} and the BLR \citep{Dehg19}.

\begin{figure}[!tbp]
\hspace{-0.6cm}\resizebox{1.13\hsize}{!}{\includegraphics[angle=0]{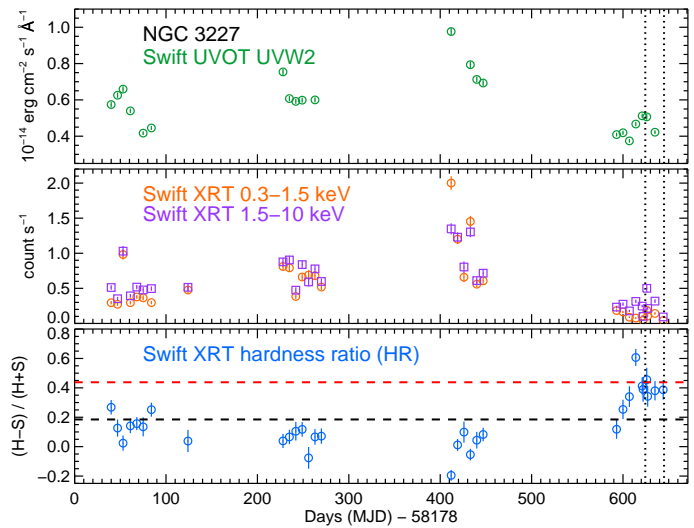}}
\caption{\swift light curve of \ngc from March 2018 to end of 2019. The dashed black line in the bottom panel indicates the average quiescent hardness ratio (HR) from unobscured
data. The dashed line in red is the HR limit for triggering, above which significant obscuration was predicted according to our simulations. The hardness ratio is defined as $(H-S)/(H+S)$, where $H$ and $S$ are the \swift XRT count rate fluxes in the hard (1.5--10 keV) and soft (0.3--1.5 keV) bands, respectively. The first and second \xmm observations, taken on 15 November and 5 December 2019, are indicated by vertical dotted lines.}
\label{swift_lc}
\end{figure}

\begin{figure}[!tbp]
\centering
\resizebox{\hsize}{!}{\includegraphics[angle=270]{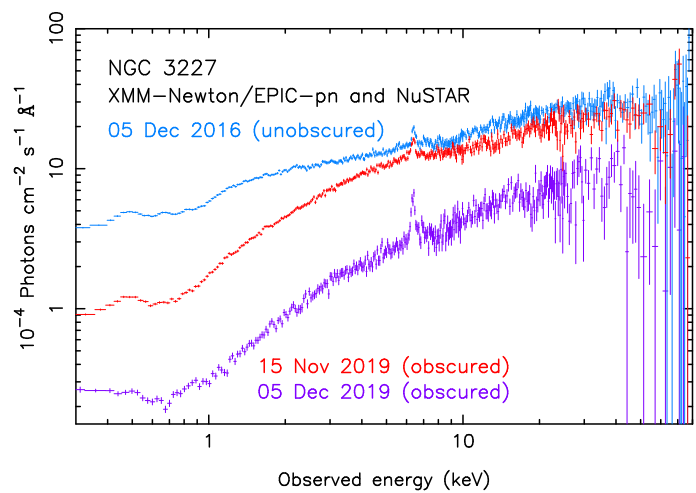}\hspace{-0.0cm}}
\caption{\ngc unobscured (2016) and obscured (2019) spectra from \xmm EPIC-pn and \nustar. The displayed energy for EPIC-pn is 0.3--10 keV and for \nustar 10--80 keV.}
\label{unobscured_obscured_fig}
\end{figure}

Following our findings in \object{NGC~3783}, we continued our \swift monitoring to find more of these events in other AGN to broaden our understanding of this transient obscuration phenomenon. In 2019 we discovered another obscuration event in the AGN \object{NGC~3227}. Figure \ref{swift_lc} shows the \swift light curve of \ngc from March 2018 to end of 2019. In November 2019, we found an intense X-ray spectral hardening event that lasted for a few weeks until the end of the \swift visibility window. During this period we successfully executed the triggering of our \xmm, \nustar, and \hst/COS observations. Two sets of observations were taken, separated by about three weeks. Figure \ref{unobscured_obscured_fig} compares an archival unobscured spectrum of \ngc taken on 5 December 2016 with the new obscured spectra taken on 15 November and 5 December 2019. Strong soft X-ray absorption is evident in the new spectra. There is also intrinsic X-ray variability between the two 2019 observations.

%

%

\ngc is a Seyfert-1 galaxy at redshift of 0.003859 \citep{deVa91} based on \ion{H}{i} 21-cm line measurements. From an \xmm and \nustar campaign carried out in 2016, \citet{Turn18} found occultation by gas clouds that cause rapid X-ray flux and spectral variability. They concluded that the occulting gas is likely associated with clouds in the inner BLR. The X-ray variability and lags seen during the campaign were further studied by \citet{Lobb20}. The spiral galaxy of \ngc is interacting with its companion galaxy \object{NGC~3226}, which is a dwarf elliptical galaxy. The \ion{H}{i} radio mapping studies of \ngc show that there is a significant column density of neutral gas in the host galaxy of \ngc as a result of tidal interactions between the two galaxies \citep{Mund95b}. Significant internal reddening is also evident from the optical-UV spectrum of \ngc \citep{Cren01a}.

In this paper (Paper I) we determine the intrinsic broadband continuum of the spectral energy distribution (SED) in \ngc, while in our following papers we study the ionised outflows (Paper II by \citealt{Wang21}), investigate the new obscuring wind in detail (Paper III by \citealt{Mao21}), and analyse the nature of variability in \ngc (Paper IV by \citealt{Graf21}). Establishing a broadband continuum model is important for (1) correctly disentangling the spectral components of obscuration from the continuum; (2) photoionisation modelling of the ionised AGN outflows, and correctly deriving and interpreting the AGN wind parameters; (3) understanding how accretion leads to the production of the observed broadband continuum emission. In order to derive the continuum, all foreground effects in our line of sight to the central engine of the AGN, such as reddening and absorption, which are prominent in \ngc, need to be properly taken into account. This means broadband continuum modelling and absorption/reddening modelling are dependent on each other.

The structure of the paper is as follows. The observations and the reduction of data are described in Sect. \ref{data_sect}. The spectral analysis and modelling of the continuum, reddening, and absorption are presented in Sect. \ref{model_sect}. We discuss our results about the origin of the internal reddening and the components of the accretion-powered SED in Sect. \ref{discussion}, and give concluding remarks in Sect. \ref{conclusions}. We adopt a luminosity distance of 16.58 Mpc in our calculations with the cosmological parameters ${H_{0}=70\ \mathrm{km\ s^{-1}\ Mpc^{-1}}}$, $\Omega_{\Lambda}=0.70$, and $\Omega_{m}=0.30$. We assume proto-solar abundances of \citet{Lod09} in all our computations in this paper.

\section{Observations and data processing}
\label{data_sect}

The observation log of data used in our SED modelling are provided in Table \ref{log_table}. We model two sets of spectra: the unobscured spectra taken by \xmm and \nustar on 5 December 2016, and the obscured spectra taken by \xmm and \nustar on 5 December 2019. Hereafter we refer to them as the `unobscured 2016' and the `obscured 2019' observations. The 2016 unobscured spectra help us in better constraining the model for the intrinsic soft X-ray continuum, which is strongly absorbed in 2019. To determine the intrinsic broadband continuum, we also use the simultaneous HST/COS UV spectrum taken on 5 December 2019, as well as an archival HST Space Telescope Imaging Spectrograph (STIS) spectrum taken in 2000 in order to extend the spectral coverage to optical and near infrared (NIR) energies. As described in Sect. \ref{intro_sect} two sets of observations were taken in 2019 (Fig. \ref{unobscured_obscured_fig}). We use the second set (5 December 2019), which is sufficient for our purpose of obtaining the SED at the obscured epoch. Also, the second obscured spectrum is the one that differs the most from the unobscured 2016 spectrum (Fig. \ref{unobscured_obscured_fig}), thus can be more useful for checking differences in the SEDs of two epochs. Furthermore, the \xmm Optical Monitor (OM) grism spectrum of the first 2019 observation (15 November 2019) had contamination from another source, and thus we use the second 2019 observation (5 December 2019). We describe the processing and preparation of data from different observatories in the following.

\subsection{NIR-optical-UV data}

Our new HST observations of NGC 3227 used COS \citep{Green12} to obtain UV spectra covering the far-UV spectral range from 1067 to 1801 \AA.
We obtained four 415~s exposures at all four focal-plane positions (FP-POS) with grating G130M at a central wavelength setting of 1222 \AA; two 480~s exposures with grating G160M with central wavelength 1533 \AA; and two 564~s exposures with grating G160M with central wavelength 1623 \AA. The diverse grating settings and focal plane positions enable us to span the gaps between detector segments A and B as well as eliminate flat-field artifacts and other detector anomalies \citep{COS_IHB20}. We retrieved the COS data from the Mikulski Archive for Space Telescopes (MAST) as processed with the latest calibration pipeline, v3.3.10. 

In order to fully determine the shape of the NIR-optical-UV continuum, we also retrieved archival spectra using the HST Space Telescope Imaging Spectrograph (STIS) \citep{Woodgate98}, which extends the spectral coverage to 10255 \AA. The STIS spectrum was scaled by a factor of 0.68 to match the 2019 COS flux at their overlapping energy bands. These data were also processed with the standard data reduction pipelines. The observation log of the HST COS and STIS observations, and the \xmm and \nustar observations that were used in our SED modelling, are provided in Table \ref{log_table}.

In our SED modelling we made use of \xmm OM \citep{Mas01} data. The data from OM, which was operated in the Science User Defined mode, were taken with the primary photometric filters and the optical grism. The size of the circular aperture used for our photometry was set to a diameter of 12$\arcsec$, which is the optimum aperture size based on the calibration of these instruments. For a description of the reduction of OM data, we refer to Appendix A in \citet{Meh15a} and references therein, which also applies to the data used here.

In this paper we also made use of the NIR continuum fluxes that were derived by \citet{Kish07} from archival HST Near Infrared Camera and Multi-Object Spectrometer (NICMOS) observations, taken with the F160W (16050 \AA) and F222M (22180 \AA) filters. We use these two continuum data points to compare with our SED model in the NIR band as described in Sect. \ref{uv_sect}.

\subsection{X-ray data}

For our SED modelling in the X-ray band, we fitted spectra from \xmm RGS and EPIC-pn, and \nustar. The \xmm data were processed using the Science Analysis System (SAS v18.0.0). The RGS \citep{denH01} instruments were operated in the standard Spectro+Q mode. The data were processed through the {\tt rgsproc} pipeline task; the source and background spectra were extracted and the response matrices were generated. We filtered out time intervals with background count rates $> 0.1\ \mathrm{count\ s}^{-1}$ in CCD number 9. The first-order RGS1 and RGS2 spectra were fitted simultaneously over the 6--37~\AA\ band in our spectral modelling.

The \xmm EPIC-pn instrument \citep{Stru01} was operated in the Small-Window mode with the Thin Filter. Periods of high-flaring background for EPIC-pn (exceeding 0.4 $\mathrm{count\ s}^{-1}$) were filtered out while applying the {\tt \#XMMEA\_EP} filter. We extracted a single event ({\tt PATTERN==0}), high-energy ({\tt PI>10000 \&\& PI<12000}) light curve from the event file over the entire chip to identify intervals of high-flaring background. The \xmm EPIC-pn spectra were extracted from a circular region centred on the source with a radius of $40''$. The background was extracted from a nearby source-free region of radius $40''$ on the same CCD as the source. The pileup was evaluated to be negligible. The single and double events were selected for the EPIC-pn ({\tt PATTERN <= 4}). Instrumental response matrices were generated for the spectrum using the {\tt rmfgen} and {\tt arfgen} tasks. The fitted spectral range for EPIC-pn is 0.3--10 keV.

The \nustar observations were reduced using the NuSTAR Data Analysis Software (NUSTARDAS) and CALDB calibration files of HEASoft v6.27. The data were processed with the standard pipeline script {\tt nupipeline} to produce level 1 calibrated and level 2 cleaned event files. The data from the South Atlantic Anomaly passages have been filtered out and event files were cleaned with the standard depth correction, which reduces the internal background at high energies. The source was extracted from a circular region (radius $\sim 90''$), with the background extracted from a source-free area of equal size on the same detector. Then the {\tt nuproducts} script was run to create level 3 products (spectra, lightcurves, ARF and RMF response files) for each of the two hard X-ray telescope modules (FPMA and FPMB) onboard \nustar. The spectra and corresponding response files of the two telescopes were combined for spectral modelling using the {\tt mathpha}, {\tt addrmf}, and {\tt addarf} tools of the {\tt HEASOFT} package. The fitted spectral range for \nustar is 5--78 keV.

Our spectral analysis and modelling were done using the {\tt SPEX} package \citep{Kaa96,Kaas20} v3.06 with C-statistics for fitting the X-ray spectra and $\chi^2$ fitting for the NIR-optical-UV data. The model parameter errors are given at the $1\sigma$ confidence level. The \xmm and \nustar spectra are optimally binned according to \citet{Kaas16} for fitting in \spex.

%
\begin{table}[!tbp]
\begin{minipage}[t]{\hsize}
\setlength{\extrarowheight}{3pt}
\caption{Log of the \ngc observations used in this paper for our SED modelling.}
\centering
\footnotesize
\renewcommand{\footnoterule}{}
\begin{tabular}{l | c c c}
\hline \hline
 & & Obs. date &  Length \\
Observatory & Obs. ID & yyyy-mm-dd & (ks)  \\
\hline
HST COS & 15673	& 2019-12-05 & 3.7 \\
HST STIS & 8479	& 2000-02-08 & 4.0 \\
\hline
\xmm & 0782520601	& 2016-12-05 & 85 \\
\xmm & 0844341401	& 2019-12-05 & 51 \\
\hline
\nustar & 60202002010	& 2016-12-05 & 41 \\
\nustar & 80502609004	& 2019-12-05 & 28 \\
\hline
\end{tabular}
\end{minipage}
\tablefoot{
The dates correspond to the start time of the observations. The HST COS observation was taken with the G130M and G160M gratings, and the STIS observation with the G140L, G230L, G430L, and G750L gratings. 
}
\label{log_table}
\end{table}

\section{Spectral analysis and modelling of the spectral energy distribution}
\label{model_sect}

Here we present our modelling of the spectral components that form the observed SED in \ngc. We jointly model the spectra of the 2016 (unobscured) and 2019 (obscured) observations. We derive the intrinsic NIR-optical-UV-X-ray continuum by modelling all the reddening and X-ray absorption in our line of sight towards the nucleus of \ngc. Our spectral modelling is not affected by the companion galaxy NGC~3226 as it is far enough from \ngc.

\subsection{NIR-optical-UV continuum and reddening}
\label{uv_sect}

\begin{figure}[!tbp]
\centering
\resizebox{\hsize}{!}{\includegraphics[angle=0]{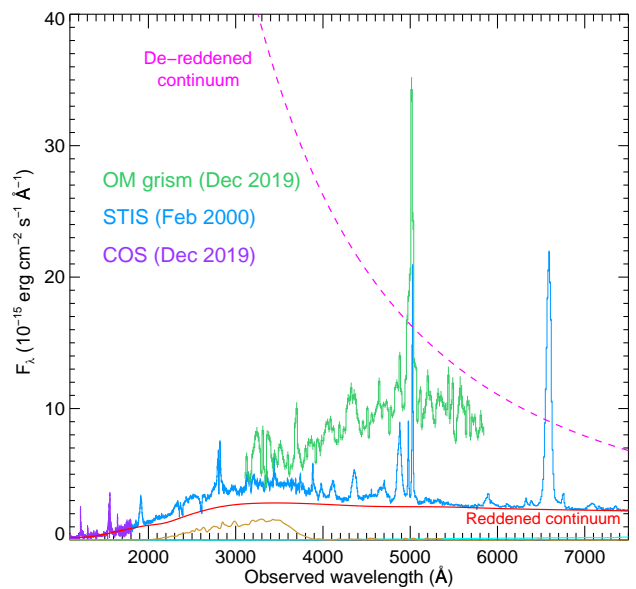}\hspace{-0.0cm}}
\caption{Overview of the reddened NIR-optical-UV spectrum of \ngc, taken with HST COS and STIS, and \xmm OM. The best-fit reddened disk blackbody model ({\tt dbb}) is shown in solid red line. For comparison the {\tt dbb} model without the effect of reddening is shown in dashed magenta line. The model component with a broad feature shown in solid dark yellow line is the blended \ion{Fe}{ii} and Balmer emission. The strong excess emission in the OM optical grism spectrum relative to the STIS spectrum is the host galaxy emission captured by the larger OM aperture. The host galaxy emission taken by the HST aperture is negligible, shown by the model component in cyan.}
\label{opt_uv_spec}
\end{figure}

The NIR-optical-UV spectrum of \ngc is shown in Fig. \ref{opt_uv_spec}. To fit the continuum emission extending from NIR to UV we started with a disk blackbody model ({\tt dbb} in \spex). This model is based on a geometrically thin, optically thick, Shakura-Sunyaev accretion disk model \citep{Sha73}. This model is appropriate for the accretion disk of the Seyfert-1 galaxy \ngc and fits well the continuum of HST COS and STIS spectra over 1067--10255 \AA, after including in the model the effect of reddening described below.

The 2016 unobscured observation does not have any HST data and its \xmm OM exposure was taken only with the UVW1 filter. Therefore, in our SED modelling we assume the shape of the NIR-optical-UV continuum in the 2016 observation was the same as in the 2019 one, and simply scale the normalisation of the {\tt dbb} model (derived from the 2019 observation) to match the 2016 OM flux. The UVW1 flux in the 2016 observation was higher by a factor of 1.82 than in the 2019 observation, thus we use this flux ratio for scaling the {\tt dbb} model.

\begin{figure*}[!tbp]
\hspace{-0.35cm}\resizebox{1.033\hsize}{!}{\includegraphics[angle=0]{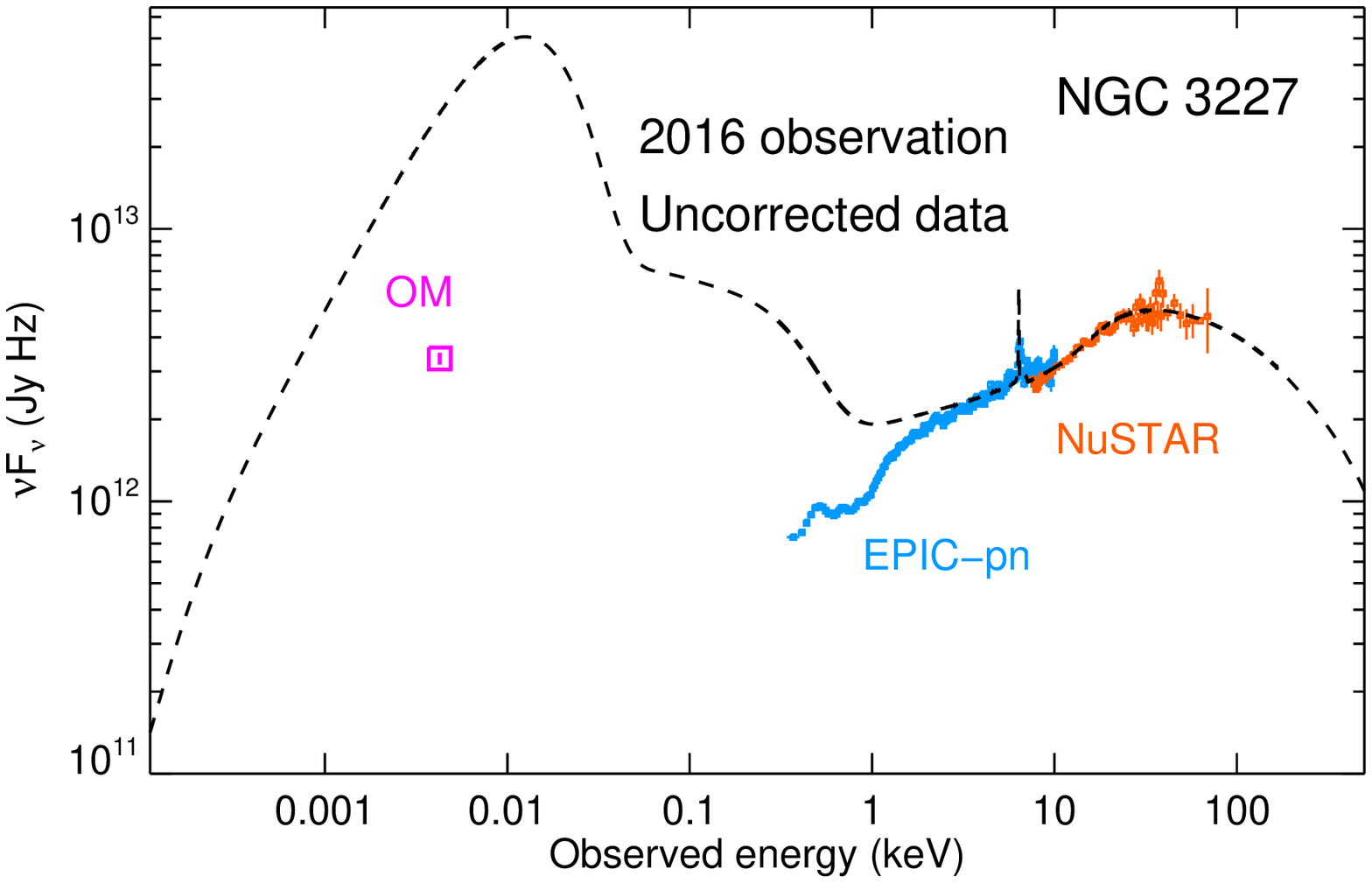}\hspace{-1cm}\includegraphics[angle=0]{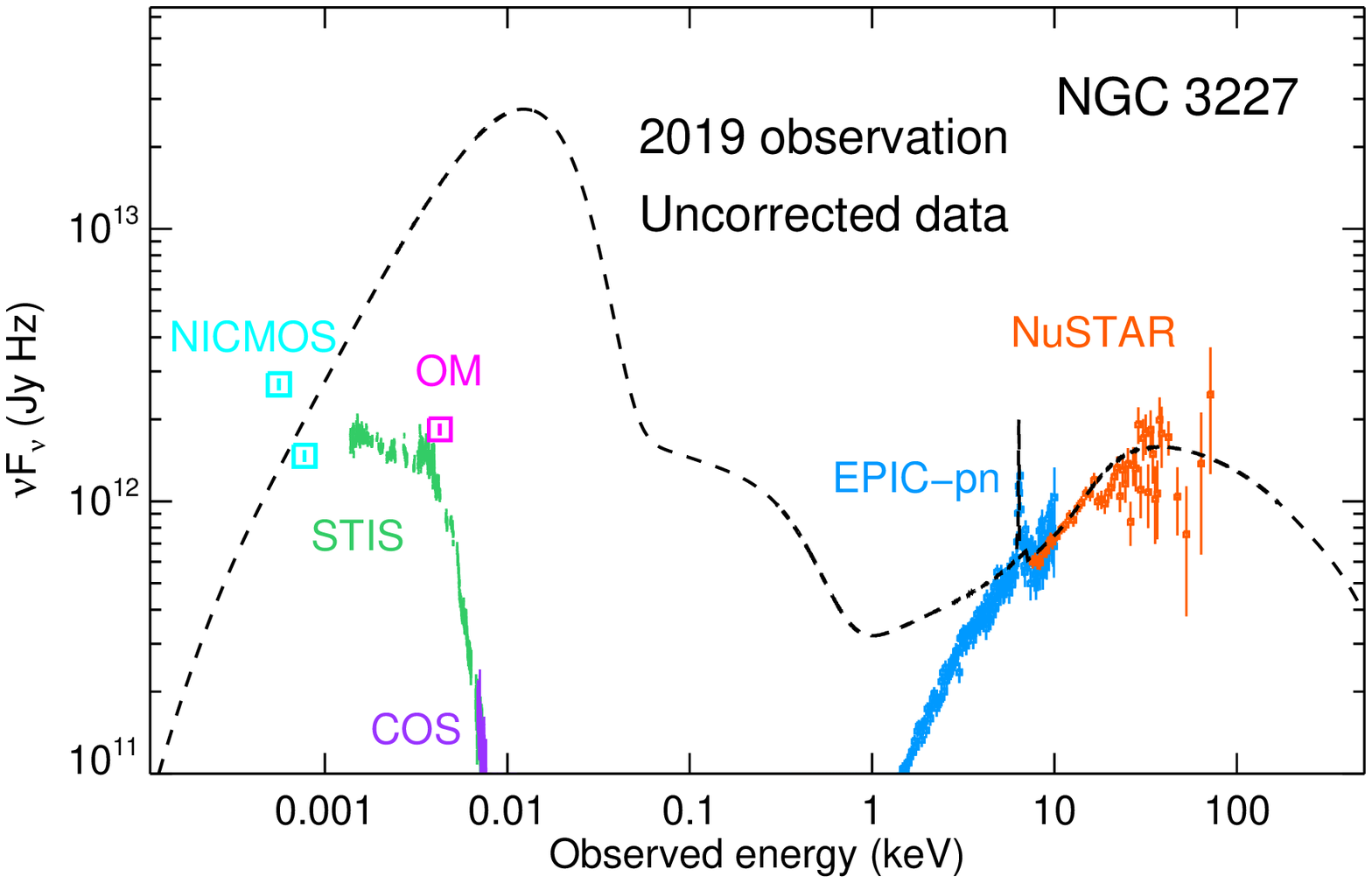}}
\vspace{-0.76cm}
\hspace{-0.35cm}\resizebox{1.033\hsize}{!}{\includegraphics[angle=0]{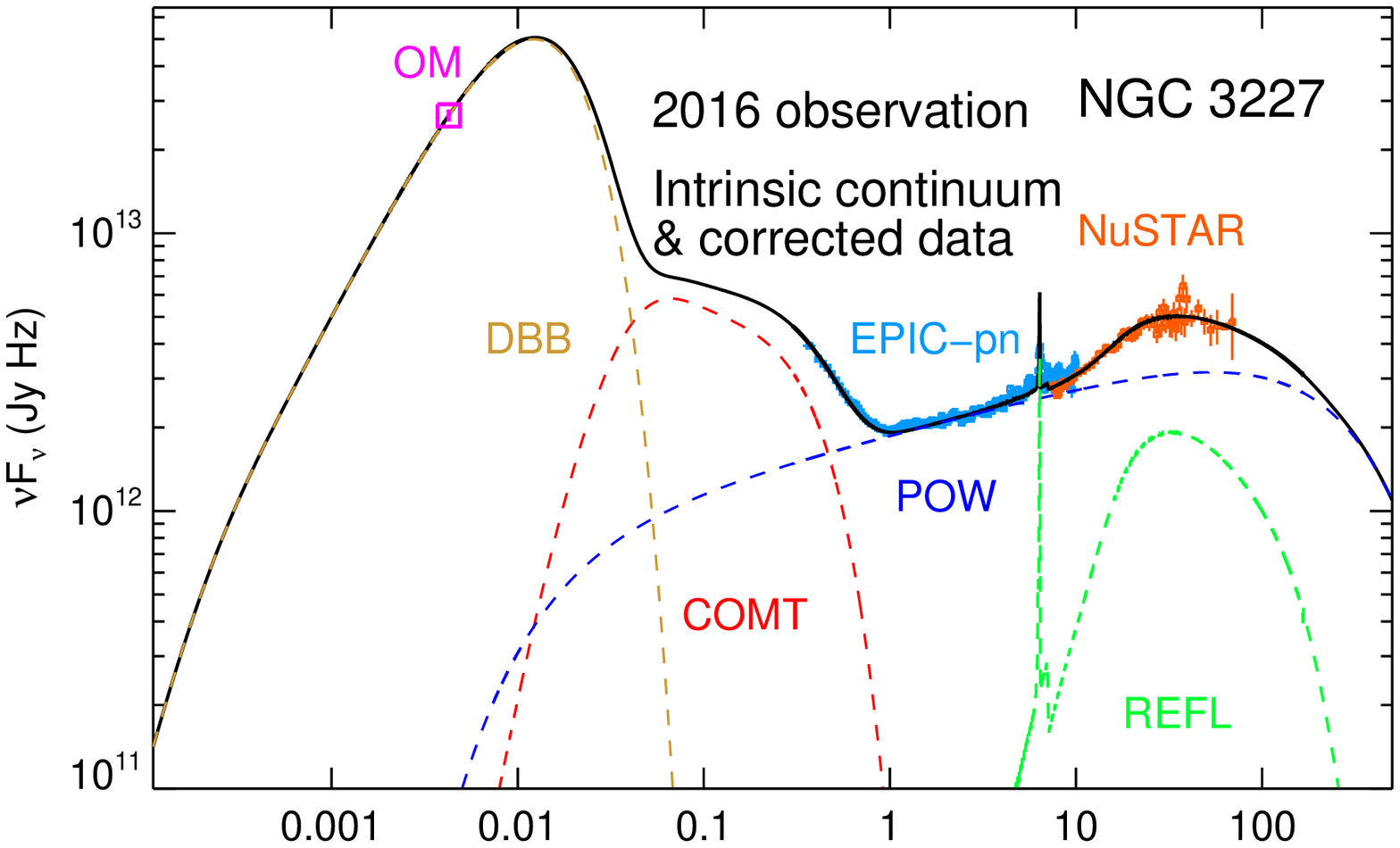}\hspace{-1cm}\includegraphics[angle=0]{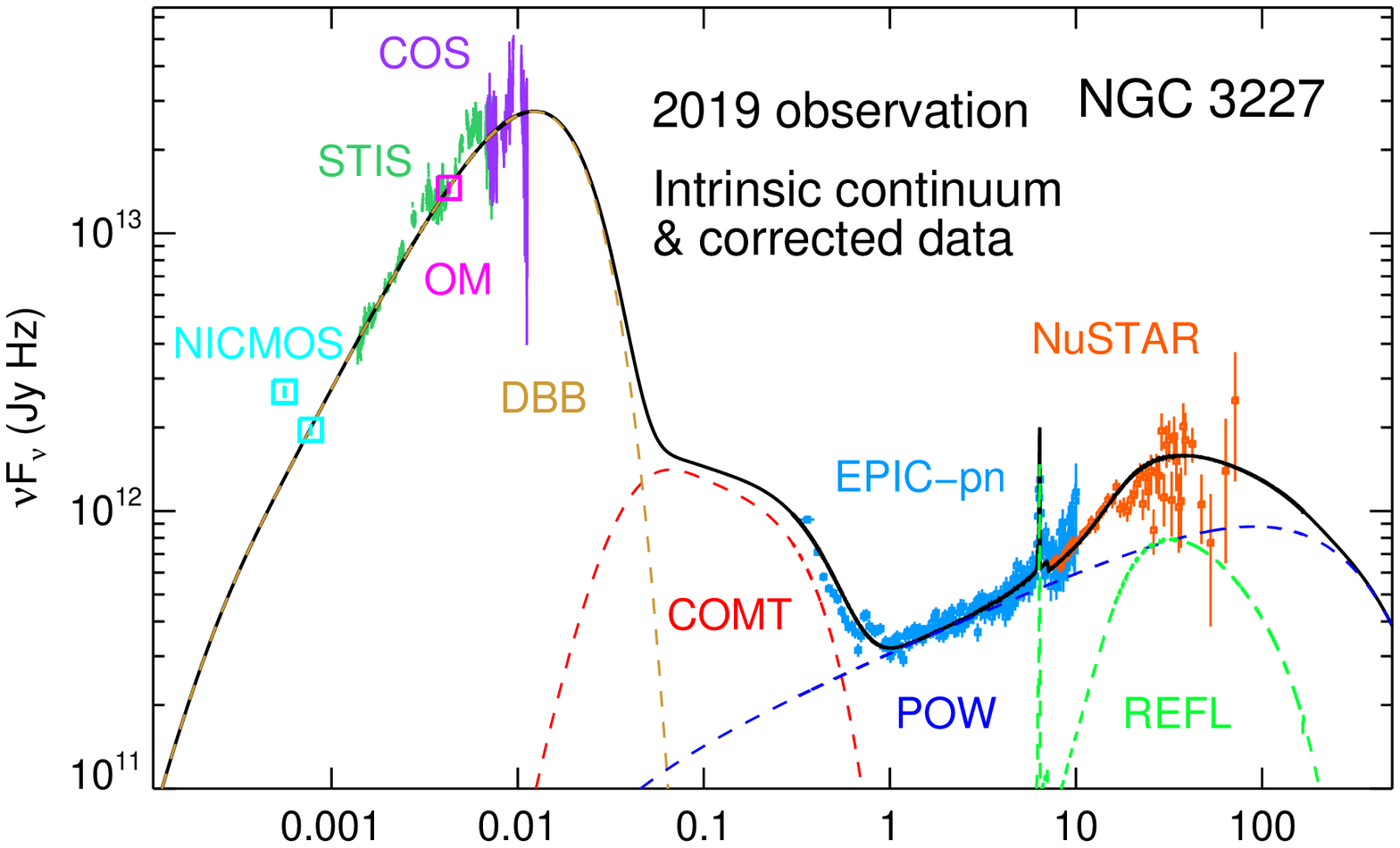}}

\hspace{-0.35cm}\resizebox{1.033\hsize}{!}{\includegraphics[angle=0]{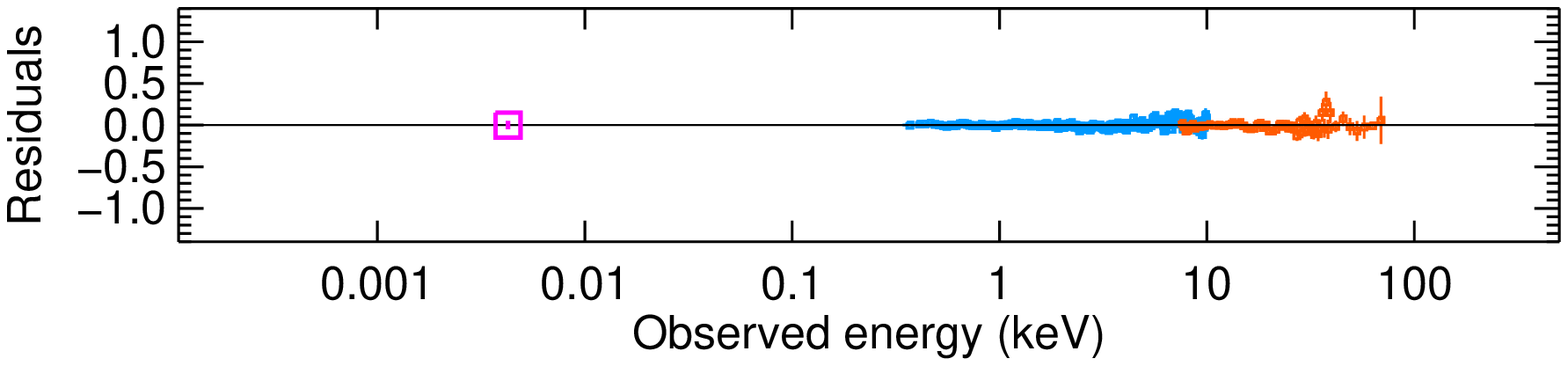}\hspace{-1cm}\includegraphics[angle=0]{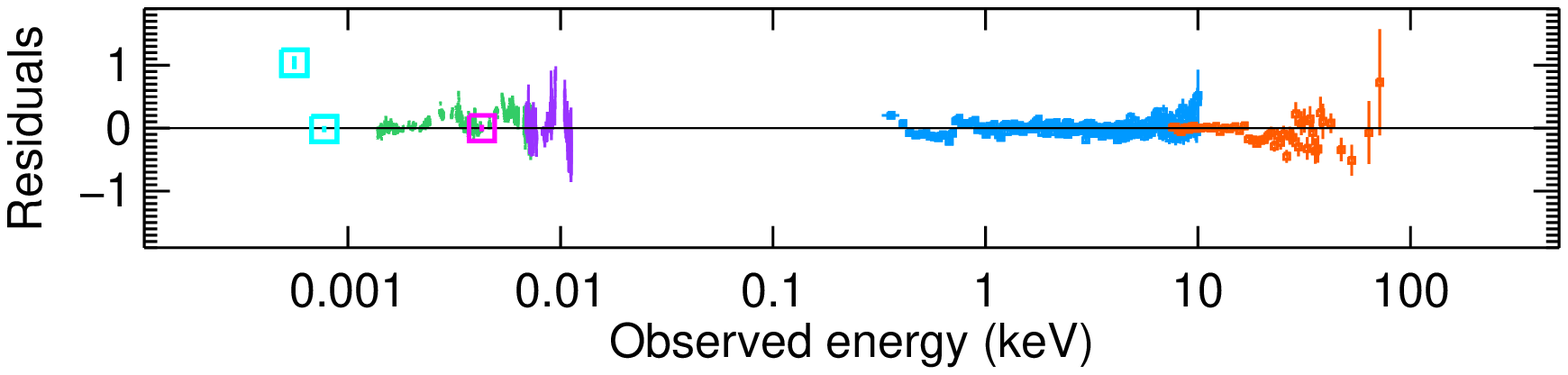}}
\caption{
SED continuum model of \ngc from NIR to hard X-rays, derived from fitting the \xmm, \nustar and HST data. The data are from the 2016 unobscured ({\it left panels}) and 2019 obscured ({\it right panels}) observations. The data in the {\it top panels} include the effects of reddening and X-ray absorption, while those in the {\it middle panels} are corrected for these effects, revealing the underlying continuum. The best-fit continuum model is shown in dashed black line in the {\it top panels} and solid black line the {\it middle panels} for comparison with the data. Residuals of the fit, defined as (data$-$model)/model, are displayed in the {\it bottom panels}. The archival HST NICMOS data at NIR are not included in our spectral fitting, but rather are over-plotted for comparison with our SED model as described in Sect. \ref{uv_sect}. The displayed spectra have been binned for clarity of presentation. The displayed OM data are taken with the UVW1 filter. The contribution of individual continuum components is displayed in the {\it middle panels}: a disk blackbody component ({\tt dbb}, dashed brown line), a warm Comptonisation component for the `soft X-ray excess' ({\tt comt}, dashed red line), a power-law continuum for the hard X-rays ({\tt pow}, dashed blue line), an X-ray reflection component ({\tt refl}, dashed green line).
}
\label{SED_fig}
\end{figure*}

\ngc displays significant internal reddening. This is evident from both the shape of the observed optical-UV continuum and the flux ratio of the AGN emission lines. As shown in the optical-UV spectrum of Fig. \ref{opt_uv_spec}, and later in the SED plot of Fig. \ref{SED_fig} (top panel), the observed optical-UV continuum drops steeply towards higher energies. This is characteristic of strong reddening. The Milky Way reddening in our line of sight is too small to account for this, as it has a colour excess ${E(B-V) = 0.02}$~mag \citep{Schl11}, reported by NASA/IPAC Extragalactic Database (NED). Therefore, the additional reddening in our line of sight is intrinsic to \ngc.

%

Before modelling the internal reddening in \ngc, we first fixed our model for the Milky Way reddening. We applied an {\tt ebv} component in \spex to model this reddening, which incorporates the extinction curve of \citet{Car89}, including the update for near-UV given by \citet{ODo94}. The ${E(B-V)}$ was set to 0.02 and the scalar specifying the ratio of total to selective extinction ${R_V = A_V/E(B-V)}$ was fixed to 3.1.

\begin{figure}[!tbp]
\hspace{-0.3cm}\resizebox{1.07\hsize}{!}{\includegraphics[angle=270]{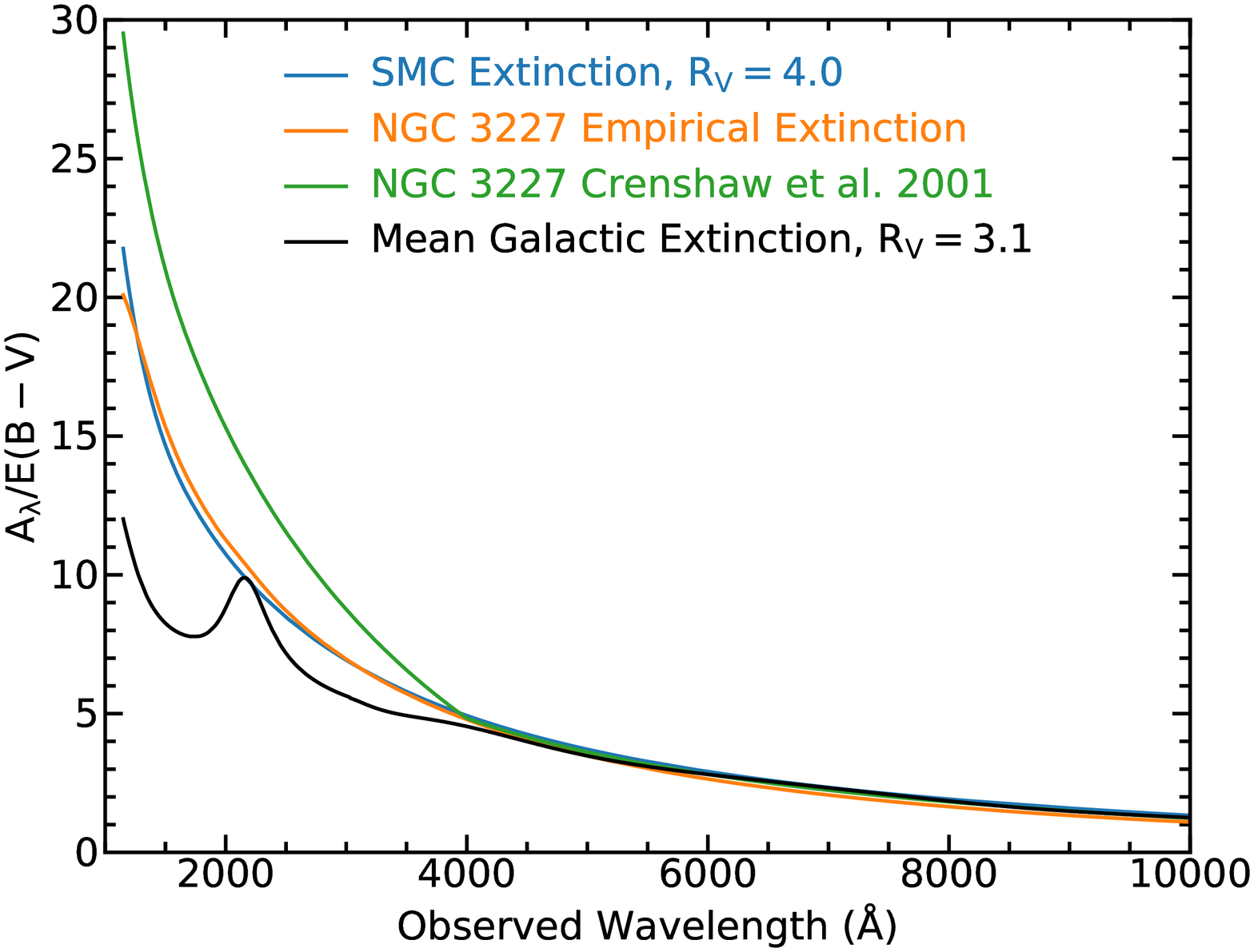}\hspace{-0.0cm}}
\caption{Comparison of different dust extinction laws which were considered in our modelling of the optical-UV spectrum of \ngc. For modelling the internal reddening in the AGN with ${E(B - V) = 0.45}$, we use the SMC extinction law of \citet{Gord03} with $R_V = 4.0$ (shown in blue). For foreground Milky Way extinction with ${E(B - V) = 0.02}$, we use the extinction curve of \citet{Car89}, including the update for near-UV given by \citet{ODo94} (shown in black).}
\label{ext_fig}
\end{figure}

To model the internal reddening in \ngc, we considered different extinction laws, which are shown in Fig. \ref{ext_fig}. They are: (1) the Milky Way extinction law (i.e. the {\tt ebv} model); (2) the extinction curve of \citet{Cren01a} derived for \ngc; (3) the Small Magellanic Cloud (SMC) extinction curve of \citet{Gord03} (the `SMC Bar Average' version) with $R_V = 4.0$; (4) an empirical extinction curve that we determined from our HST COS and STIS spectra, assuming that the intrinsic spectrum is a power law with $f_\lambda \propto \lambda^{-7/3}$, which corresponds to the Jeans tail of a Shakura-Sunyaev disk spectrum.

\citet{Cren01a} derived their extinction curve for \ngc by comparison to \object{NGC~4151}, which they assumed had no internal extinction. As shown in Fig. \ref{ext_fig}, their curve and our empirical curve are both significantly steeper than the mean Milky Way extinction law and neither include a `2175 \AA\ bump'. Therefore, the Galactic extinction curve is not suitable for correcting the internal reddening of \ngc. However, our empirically determined extinction curve is not as steep as that of \citet{Cren01a}. Instead, it closely matches an SMC extinction curve with $R_V = 4.0$ (Fig. \ref{ext_fig}). Therefore, in our SED modelling we adopted this SMC extinction curve to correct for internal reddening in \ngc. We note that internal extinction in AGN appears to be best described in general by an SMC extinction curve \citep{Hopk04}.

%

The user-defined multiplicative model in \spex ({\tt musr}) was used to import the SMC extinction model into \spex. From {\tt dbb} continuum modelling of the HST COS and STIS spectra, as well as the OM data, we find ${E(B-V) = 0.45\pm 0.03}$ best fits the shape of the observed NIR-optical-UV data. The best-fit parameters of the {\tt dbb} model are provided in Table \ref{continuum_table}. In our fitting of the continuum, we exclude regions in spectra where emission and absorption lines are present. In our modelling we included a template model component for the starlight emission from the bulge of the host galaxy, taken from \citet{Kin96}, and allowed the normalisation of this component to be fitted. As the OM data in the optical band are strongly dominated by the host galaxy starlight emission due to the larger aperture size than HST (see Fig. \ref{opt_uv_spec}), we only use the STIS spectrum for modelling the continuum in the optical band. We find STIS takes in a negligible amount of the starlight emission (Fig. \ref{opt_uv_spec}), and thus it better facilitates modelling of the underlying optical continuum than OM. In our SED modelling we included components to take into account contributions from \ion{Fe}{ii} and Balmer continuum emission in \ngc (see Fig. \ref{opt_uv_spec}). We refer to \citet{Meh15a} where these models are described and applied to NGC~5548.

\begin{figure}[!tbp]
\hspace{-0.25cm}\resizebox{1.07\hsize}{!}{\includegraphics[angle=270]{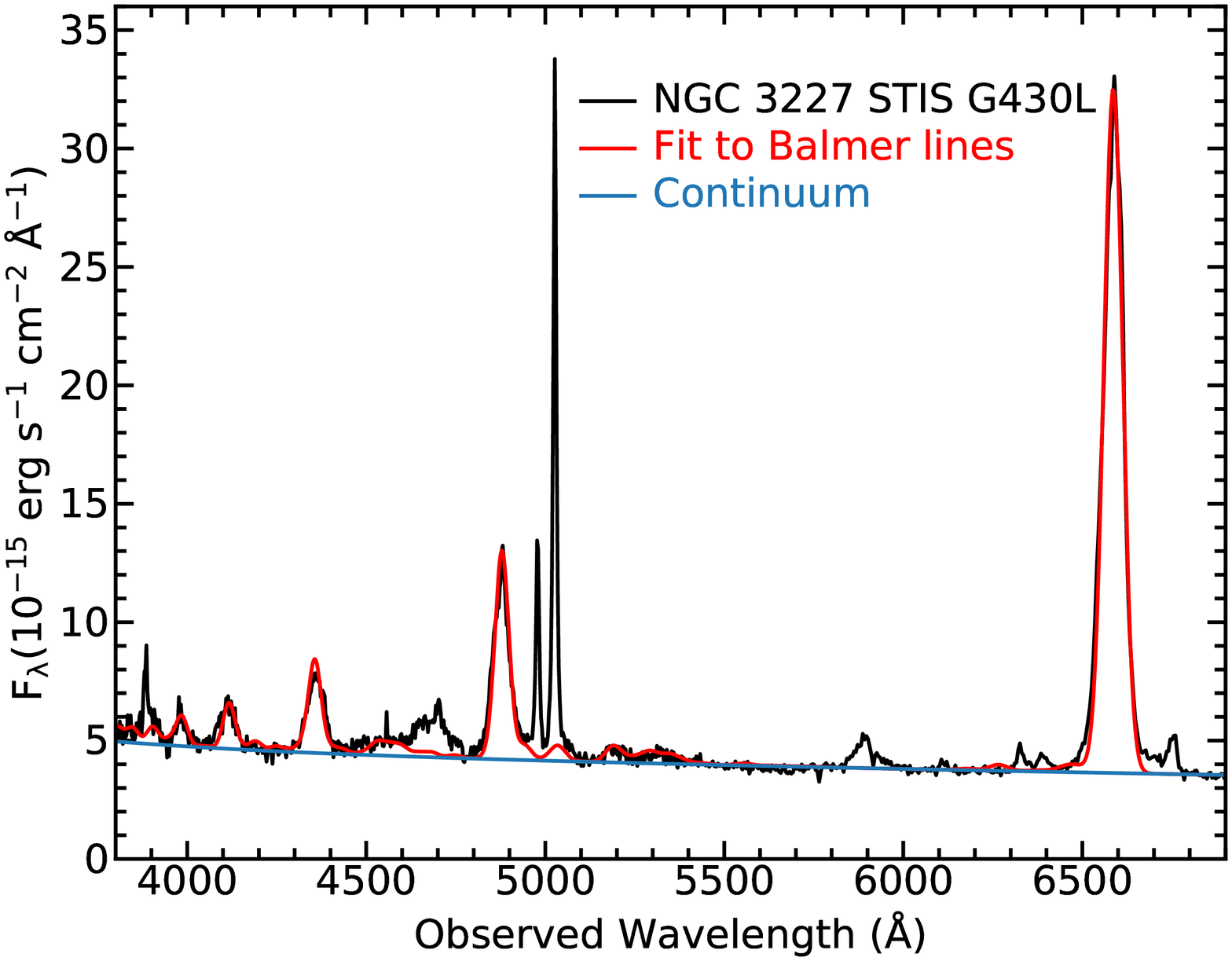}\hspace{-0.0cm}}
\caption{Close-up of the STIS spectrum of \ngc displayed in the observed frame, where the Balmer lines intensities have been fitted using a \cloudy BLR model, reddened with an SMC extinction law with $R_V=4.0$ and best-fit reddening of ${E(B-V)=0.44}$.}
\label{Balmer_fig}
\end{figure}

The internal reddening ${E(B-V) = 0.45}$ derived from modelling the continuum is fully consistent with the reddening derived from the intensity of the Balmer lines in \ngc (Fig. \ref{Balmer_fig}). Several studies have shown that the Balmer line ratios in the BLR of AGN are intrinsically unreddened and have an intensity ratio for H$\alpha$/H$\beta$ comparable to the Case B value of 2.78 \citep{Bake38} for a temperature of $10^4$~K \citep{Dong08,Gask17}. The mean unreddened value is 2.72 with an observational dispersion of ${\sim 0.3}$ \citep{Gask17}. To fit the full range of Balmer lines in the NGC~3227 spectrum, we use a typical BLR model calculated with \cloudy v17.00 \citep{Ferl17}, using the unobscured spectral energy distribution of NGC ~5548 (see \citealt{Stee05}) with an ionisation parameter $U$ \citep{Davi77} of ${\log U = -1.08}$, density ${n_{\rm H} = 10^{10}}$~cm$^{-3}$, and a total column density of ${N_{\rm H} = 1.20 \times 10^{22}}$~cm$^{-2}$. This model gives an intrinsic (unreddened) H$\alpha$/H$\beta$ intensity ratio of 2.58. In Fig. \ref{Balmer_fig} the fitted components are a power-law continuum, \ion{Fe}{ii} emission, and ${E(B-V)}$ for an SMC extinction law with $R_V=4.0$. The best fit we obtain for the line ratios is ${E(B-V) = 0.44}$. Allowing for the observed scatter in AGN Balmer line ratios, our fitted extinction has an uncertainty of $\pm$0.05. Our best-fit value is remarkably almost the same as the value obtained from modelling the shape of the full observed optical-UV continuum described earlier. So the derived $E(B-V)$ based on the lines alone in the optical region are consistent with the same extinction law used to model the whole continuum spectrum from NIR to far-UV. In Sect. \ref{ext_discuss} we will further discuss the nature and origin of internal reddening in \ngc.

%

\begin{figure}[!tbp]
\hspace{-0.2cm}\resizebox{1.04\hsize}{!}{\includegraphics[angle=0]{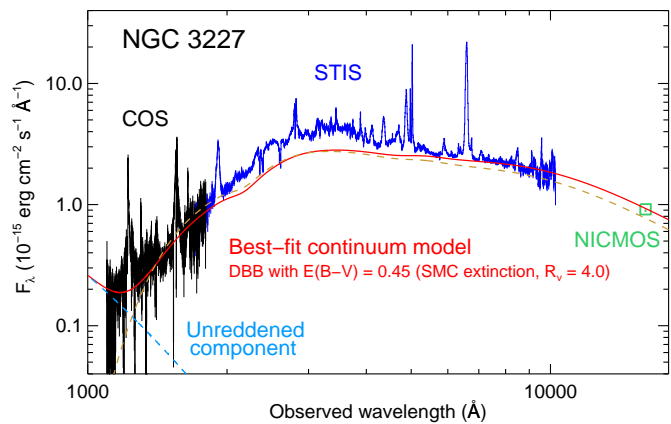}\hspace{-0.0cm}}
\caption{NIR-optical-UV spectrum of \ngc taken with HST COS and STIS. The continuum flux in the NIR from HST/NICMOS is also plotted. The best-fit continuum model, consisting of a reddened {\tt dbb} and an additional unreddened component at shortest wavelengths (dashed blue line) is shown as a solid red line. For comparison the best-fit model without an unreddened component is also plotted (dashed brown line), which does not fit the data well at shortest wavelengths. The broad feature above the continuum at about 2000--4000~\AA\ is the blended \ion{Fe}{ii} and Balmer emission.}
\label{unred_fig}
\end{figure}

Interestingly, apart from the reddened {\tt dbb} component, we also find evidence of an unreddened component. This component is seen as an excess emission above the {\tt dbb} model towards the shortest wavelengths in the COS band (see Fig. \ref{unred_fig}). As shown in this figure by including an unreddened component at the far-UV we obtain a significantly better fit to the NIR-optical-UV spectrum. As the unreddened component is weak and only observable over a limited energy band in the far-UV, distinguishing between different possible models for this emission would be challenging. Over 1067--1300 \AA, where this component is detected in the COS spectrum, its flux is about $3.4 \times 10^{-14}$ \ergflux. This is about a factor of three larger than the reddened {\tt dbb} flux. However, its flux is about three orders of magnitude smaller than the intrinsic {\tt dbb} flux. As the flux of this unreddened component increases towards higher energies, it is likely it gets stronger in the extreme ultraviolet (EUV) band. To test this and estimate a maximum flux contribution for this component, we applied an unreddened version of the warm Comptonisation model that is used to model the EUV continuum and the `soft X-ray excess' (see the {\tt comt} modelling described in Sect. \ref{xray_sect}). The inclusion of this unreddened component improves the fit by ${\Delta \chi^2 = 1300}$. We find such an unreddened {\tt comt} model, which fits the excess emission in the COS band well, has a flux that is still about two orders of magnitude smaller than the intrinsic SED model in the EUV band. Therefore, this unreddened component, which likely originates far from the central ionising source, is much weaker than the primary SED continuum and thus for the purpose of photoionisation it is effectively negligible. In Sect. \ref{ext_discuss} we will further discuss the possible origin and location of this unreddened component that is seen in the COS spectrum of \ngc.

%

In order to examine how our {\tt dbb} continuum model extends to lower NIR energies beyond the STIS band, we made use of continuum fluxes that were derived by \citet{Kish07} using archival HST Near Infrared Camera and Multi-Object Spectrometer (NICMOS) observations, taken with the F160W (16050 \AA) and F222M (22180 \AA) filters on 6 April 1998. These data are not included in our spectral fitting, but rather are over-plotted for comparison with our SED model. As shown in Figs. \ref{unred_fig} and \ref{SED_fig} (right middle panel), the measured continuum at 16050 \AA\ is consistent with our {\tt dbb} model. However, at the longer wavelength of 22180 \AA, the flux of the measured continuum is higher than that of the {\tt dbb} model (see Fig. \ref{SED_fig}, right middle panel). This excess above the {\tt dbb} model is likely the high-energy tail of a thermal emission component from the hottest regions of the AGN torus. Indeed by going to longer wavelengths the SED becomes dominated by emission from the torus rather than the accretion disk, as shown for example in \citet{Meh18b} for IC~4329A. Since in the current paper we are interested in the primary emission from the accretion disk and the associated X-ray emission, we do not extend our spectral coverage to lower energies in the mid-IR to model the torus emission.

%
\begin{table}[!tbp]
\begin{minipage}[t]{\hsize}
\setlength{\extrarowheight}{3pt}
\caption{Best-fit parameters of the broadband continuum model components for \ngc, derived from modelling the 2016 and 2019 observations.}
\centering
\small
\renewcommand{\footnoterule}{}
\begin{tabular}{l | c}
\hline \hline
Parameter						& Value					\\
\hline
\multicolumn{2}{c}{Disk blackbody component for NIR-optical-UV continuum ({\tt dbb}):}		 						\\
Normalisation $A$ &  $7.3$ (2016, fixed)			\\
                  &  $4.0 \pm 0.2$ (2019)		\\
$T_{\rm max}$ (eV)					& ${10.2}$ (2016, coupled)			\\
                  					& ${10.2 \pm 0.2}$ (2019)		  \\
\hline
\multicolumn{2}{c}{Warm Comptonisation component for `soft X-ray excess' ({\tt comt}):} 						\\
Normalisation	&		${5.9 \pm 0.1}$ (2016) \\
             	&		${1.3 \pm 0.1}$ (2019) \\
$T_{\rm seed}$ (eV) &	${10.2}$ (coupled) \\
$T_{\rm e}$ (keV) &		${0.10 \pm 0.01}$ (2016) \\
                 &		${0.10}$ (2019, coupled) \\
Optical depth $\tau$ &	${30}$ (fixed) \\
\hline
\multicolumn{2}{c}{Primary X-ray power-law component ({\tt pow}):} 						\\
Normalisation			& ${38.7 \pm 0.1}$ (2016) \\
            			& ${6.4 \pm 0.1}$ (2019) \\
Photon index $\Gamma$		& ${1.83 \pm 0.02}$	(2016)		\\
                    		& ${1.70 \pm 0.03}$	(2019)		\\
\hline
\multicolumn{2}{c}{X-ray reflection component ({\tt refl}):} 						\\
Incident power-law Norm.			& ${38.7}$ (coupled) \\
Incident power-law $\Gamma$		& ${1.83}$ (coupled)		 			\\
Reflection scale	$s$			& $0.61 \pm 0.02$ (2016)					\\
                	  			& $0.25 \pm 0.02$ (2019)					\\
\hline
\multicolumn{2}{c}{C-stat\,/ expected C-stat = 4085\,/\,3272 (2016)}  \\
\multicolumn{2}{c}{C-stat\,/ expected C-stat = 3891\,/\,3487 (2019)}  \\
\hline
\end{tabular}
\end{minipage}
\tablefoot{
The disk blackbody {\tt dbb} normalisation is in $10^{26}$ cm$^{2}$. The normalisation of the 2016 {\tt dbb} component is fixed at a factor of 1.82 larger than that of the 2019 {\tt dbb} component based on the observed OM UVW1 fluxes. The power-law normalisation of the {\tt pow} and {\tt refl} components is in units of $10^{49}$ photons~s$^{-1}$~keV$^{-1}$ at 1 keV. The normalisation of the Comptonisation component ({\tt comt}) is in units of $10^{53}$ photons~s$^{-1}$~keV$^{-1}$. The high-energy exponential cut-off of the power-law for both {\tt pow} and {\tt refl} is fixed to 309~keV. The seed photon temperature $T_{\rm seed}$ of {\tt comt} is coupled to the maximum temperature $T_{\rm max}$ of {\tt dbb}. The normalisation of the incident power-law for the reflection component ({\tt refl}) is coupled to that of the observed primary power-law continuum ({\tt pow}). The intrinsic bolometric luminosity of each continuum component is as follows. For the 2016 observation: {\tt dbb}: ${3.4 \times 10^{43}}$~\ergs; {\tt comt}: ${5.1 \times 10^{42}}$~\ergs; {\tt pow}: ${7.1 \times 10^{42}}$~\ergs; {\tt refl}: ${1.3 \times 10^{42}}$~\ergs. For the 2019 observation: {\tt dbb}: ${1.9 \times 10^{43}}$~\ergs; {\tt comt}: ${1.1 \times 10^{42}}$~\ergs; {\tt pow}: ${1.6 \times 10^{42}}$~\ergs; {\tt refl}: ${5.4 \times 10^{41}}$~\ergs. 
}
\label{continuum_table}
\end{table}

%

\subsection{X-ray continuum and absorption}
\label{xray_sect}

We simultaneously model \xmm EPIC-pn, RGS, and \nustar spectra of each observation. We took into account the instrumental flux calibration differences by comparing the flux level of the instruments at their overlapping energy bands and re-scaling their cross-normalisation. The \nustar/EPIC-pn flux ratio was 1.09 in 2016 and 1.11 in 2019; the RGS/EPIC-pn flux ratio was 0.96 in 2016 and 0.91 in 2019. To fit the X-ray spectra of \ngc we started with a power-law continuum ({\tt pow}). The X-ray power-law model represents Compton up-scattering of the disk photons in an optically-thin, hot, corona. The high-energy exponential cut-off of the power-law was first set to 309 keV \citep{Turn18}, which is consistent with the \nustar data. We then tested freeing this parameter in our modelling, however, since it did not significantly improve the fit we kept it frozen at 309~keV to limit the number of free parameters. A low-energy exponential cut-off was also applied to the power-law continuum to prevent it exceeding the maximum energy of the seed {\tt dbb} disk photons (10~eV). The photon index $\Gamma$ of the intrinsic power-law is ${1.83 \pm 0.02}$ in 2016 and ${1.70 \pm 0.03}$ in 2019.

In addition to the power-law, the soft X-ray continuum of \ngc shows the presence of a `soft X-ray excess' component (see e.g. \citealt{Mark09,Noda14}). To model the soft excess in \ngc, we used a warm Comptonisation model ({\tt comt} in \spex), which has been used to model the soft excess in similar Seyfert-1 AGN, such as \object{Mrk~509} \citep{Meh11}; NGC 5548 \citep{Meh15a}; NGC~3783 \citep{Mehd17}. In this explanation of the soft excess, the seed disk photons are up-scattered in a warm, optically thick, corona to produce the EUV continuum and the soft X-ray excess as its high-energy tail. Multi-wavelength studies have found warm Comptonisation to be a viable explanation for the soft X-ray excess in Seyfert-1 AGN (e.g. \object{Ark~120}, \citealt{Porq18}). For more details about warm Comptonisation modelling of the soft X-ray excess, see e.g. \citet{Mag98, Meh11, Done12, Petr13,Petr18,Petr20,Kubo18}. We note that this is only one plausible explanation proposed in the literature for the origin of the soft X-ray excess in AGN (relativistic ionised reflection is another explanation), yet it is sufficient for our purpose of deriving a broadband SED for \ngc in this paper. The parameters of the {\tt comt} model are its normalisation, seed photon temperature $T_{\rm seed}$, electron temperature $T_{\rm e}$, and optical depth $\tau$ of the up-scattering plasma. In our modelling the $T_{\rm seed}$ parameter of {\tt comt} is coupled to the $T_{\rm max}$ parameter of the {\tt dbb} model that fits the NIR-optical-UV data (Sect. \ref{uv_sect}). Because in the 2019 observation the soft X-ray spectrum is obscured, it can be challenging to constrain the parameters of the soft X-ray excess. Thus, we couple the 2019 {\tt comt} plasma parameters to those of the unobscured 2016 observation and only allow its normalisation to be fitted. Furthermore, as there can be degeneracy between plasma parameters of the soft excess model, we fix optical depth $\tau$ of the optically-thick plasma to a fiducial value of 30, in order to limit the number of free parameters while still providing a good fit. The best-fit parameters of the power-law component ({\tt pow}), the warm Comptonisation component ({\tt comt}), and the disk blackbody model ({\tt dbb}, described in Sect. \ref{uv_sect}) are provided in Table \ref{continuum_table}. The continuum components of our SED model, from NIR to hard X-rays, are displayed in Fig. \ref{SED_fig} (middle panels).

The primary X-ray continuum undergoes reprocessing, which is evident by the presence of the \FeKa line and the Compton hump in the \xmm and \nustar spectra (Fig. \ref{SED_fig}). We applied an X-ray reflection component ({\tt refl} in \spex), which reprocesses an incident power-law continuum to fit the \FeKa line and the Compton hump at hard X-rays. The {\tt refl} model computes the \FeKa line according to \citet{Zyck94}, and the Compton-reflected continuum according to \citet{Magd95}, as described in \citet{Zyck99}. The exponential high-energy cut-off of the incident power-law was also set to that of the observed primary power-law component at 309~keV \citep{Turn18}. In our modelling the normalisation of the incident power-law continuum was coupled to that of the 2016 observed power-law. The ionisation parameter of {\tt refl} is set to zero to produce a cold reflection component with all abundances kept at their solar values. The photon index $\Gamma$ of the incident power-law is fixed to that of the primary power-law seen in the unobscured 2016 observation. The reflection scale factor ($s$) of the {\tt refl} model was fitted for the 2016 and 2019 observations. The best-fit parameters are provided in Table \ref{continuum_table}.

In our spectral modelling we take into account the X-ray continuum and line absorption by the diffuse interstellar medium (ISM) in the Milky Way. This is done using the {\tt hot} model in \spex \citep{dePl04,Stee05}. This model calculates the transmission of a plasma in collisional ionisation equilibrium at a given temperature, which for neutral ISM is set to the minimum temperature of the model at 0.008 eV. The Galactic \NH in our line of sight towards \ngc was fixed to $2.07 \times 10^{20}$~\cm \citep{Mur96}. 

Absorption by ionised AGN outflows is evident in \ngc (see. e.g. \citealt{Turn18}). In our SED modelling we take into account absorption by the persistent warm-absorber outflow, and the additional absorption by the transient obscuring wind in the new 2019 data. The X-ray spectra of \ngc do not show evidence of any detectable absorption by neutral gas in the ISM of the host galaxy. As noted in Sect. \ref{intro_sect}, the study of the ionised outflows are presented in detail separately in Paper II \citep{Wang21} for the warm absorber, and Paper III \citep{Mao21} for the new obscuring wind. Below we briefly describe how the X-ray absorption by these ionised outflows were taken into account in our SED modelling. 

For photoionisation and spectral modelling, we use the \pion model in \spex (see \citealt{Meh16b}), which is a self-consistent model that calculates both the photoionisation solution and the spectrum of a plasma in photoionisation equilibrium (PIE). The \pion model uses the SED continuum model that is fitted to the data (Fig. \ref{SED_fig}, middle panels), thus both the continuum and the absorption spectrum are fitted together. The warm-absorber model that we use is based on the model derived from archival \xmm/RGS observations of \ngc (Paper II, \citealt{Wang21}). This warm-absorber model consists of four different ionisation components (i.e. four \pion components). There are no significant soft X-ray emission lines detected in the RGS spectrum of \ngc. 

The column density \NH and the ionisation parameter $\xi$ \citep{Kro81} of the warm-absorber components of the \citet{Wang21} model were freed in our modelling of the 2016 unobscured observation to improve the fit to our data. The total \NH of the warm absorber is found to be about ${4.8 \times 10^{21}}$~\cm, distributed over four ionisation components:\\ 
(A) ${\log \xi = 3.0 \pm 0.1}$ and ${\NH = 1.0 \pm 0.1 \times 10^{21}}$~\cm; \\ (B) ${\log \xi = 2.7 \pm 0.1}$ and ${\NH = 1.2 \pm 0.2 \times 10^{21}}$~\cm; \\ (C) ${\log \xi = 1.8 \pm 0.1}$ and ${\NH = 9.7 \pm 0.9 \times 10^{20}}$~\cm; \\ (D) ${\log \xi = -1.8 \pm 0.2}$ and ${\NH = 1.6 \pm 0.1 \times 10^{21}}$~\cm. 

To model the new obscuration in the 2019 data one additional \pion component is required, which is partially covering the X-ray source. This is needed to be able to fit the characteristic curvature seen in the 2019 broadband X-ray spectrum (Figs. \ref{unobscured_obscured_fig} and \ref{SED_fig}). In our modelling of the 2019 obscured observation, the ionisation parameter $\xi$ of the above warm-absorber components were self-consistently lowered in response to the shielding of the ionising SED by the obscurer. Thus, for the 2019 observation the warm-absorber parameters are not fitted, but rather the ionisation parameter is automatically re-calculated by the {\tt pion} model for the obscured SED. The new ionisation parameter of the above components corresponds to $\log \xi$ of 2.4, 2.1, 1.2, and $-2.4$, respectively. So the 2016 warm-absorber model is `de-ionised' in the 2019 observation, thus producing stronger absorption than in 2016. Such de-ionisation of the warm absorber by the obscurer has been previously presented for the obscuration in NGC~5548 \citep{Kaas14} and NGC~3783 \citep{Mehd17}. Our fitted {\tt pion} parameters of the obscurer in \ngc are: ${\NH = 5.4 \times 10^{22}}$~\cm, ${\log \xi = 1.0}$, and covering fraction ${C_f = 0.64}$. More details about the new obscuring wind and its variability in \ngc will be presented in our follow-up Paper III \citep{Mao21} and Paper IV \citep{Graf21}. 

Finally, we tested how much our SED model changes if an ionised relativistically-blurred reflection component is added to our model, where it would contribute to the soft X-ray excess (see e.g. \citealt{Patr12,Beuc15}). For this test we added a second {\tt refl} component and allowed its ionisation parameter to be fitted with relativistic disk broadening. We found the change on the photon index of the power-law and flux of the SED in the soft and hard X-ray bands are at levels of 1 to 3\%. This amount of change, which is comparable to the uncertainty of the continuum parameters, is not significant for photoionisation modelling (see e.g. \citealt{Kara21}). Therefore, considering the above test, and lack of a consensus in previous publications of \ngc about the nature of the soft X-ray excess, we keep our SED model as it was before the test. Our aim in this paper is not distinguishing between different models for the soft excess in \ngc, but rather adopt one feasible model for producing the overall SED.

\section{Discussion}
\label{discussion}
%

\subsection{Nature and origin of dust extinction in NGC~3227}
\label{ext_discuss}

We showed in Sect. \ref{uv_sect} that the optical-UV spectra of \ngc display strong internal reddening. This is evident in both the shape of the continuum (Figs. \ref{unred_fig} and \ref{SED_fig}) and the Balmer lines (Fig. \ref{Balmer_fig}). Here we discuss the possible origin and location of this dust reddening in \ngc. The host galaxy of \ngc is interacting with its companion galaxy NGC~3226. The \ion{H}{i} radio mapping studies of \ngc show interesting tidal interactions between the two galaxies. As a result there is a significant column density of neutral gas in the host galaxy of \ngc. According to radio observations, the \ion{H}{i} 21-cm absorption suggests a neutral gas column density of $5.5 \times 10^{21}$~\cm along our line of sight \citep{Mund95b}. However, interestingly, our line of sight to the X-ray and UV source does not show the presence of any significant amount of intrinsic neutral gas as discussed below. Apart from absorption by ionised AGN outflows (Sect. \ref{xray_sect}), the X-ray spectra of \ngc do not show evidence of any additional absorption by neutral gas, such as in the ISM of \ngc. This is also supported from HST UV spectroscopy of the intrinsic Ly$\alpha$ line in \ngc, which shows strong absorption, but does not have broad damping wings. This implies a total intrinsic \ion{H}{i} column of $<10^{19}~\rm cm^{-2}$. The most feasible explanation for this discrepancy between the \ion{H}{i} absorption in radio, and lack of \ion{H}{i} absorption in X-rays and UV, is that the radio continuum emission is not coincident with the nucleus of \ngc, where the X-ray and UV point source reside. Indeed the high resolution radio observations of \citet{Mund95a} show that the radio emission is extended along the direction of the ionization cone defined by [\ion{O}{iii}] $\lambda$5007 and the extended X-ray emission \citep{Zeng03}.

Since the internal reddening ${E(B-V) = 0.45}$ cannot be associated with neutral ISM gas in \ngc, one likely explanation is that the internal reddening is linked to nuclear gas, likely the outflowing warm-absorber gas that may originate from the dusty torus of \ngc. The relation between the hydrogen column density \NH and reddening \ebv can be written as ${N_{\rm{H}}\ ({\rm{cm}}^{-2}) = a \times 10^{21}\ \ebv\ (\rm{mag})}$, where $a$ is reported in the literature to have a value ranging from about 5.5 to 6.9 \citep{Bohl78,Gor75,Pre95,Guv09}. Therefore, for our derived ${\ebv = 0.45}$, the expected \NH is about ${3 \times 10^{21}}$~\cm. Interestingly, this \NH matches the column density \NH of the lower-ionisation components of the warm absorber in \ngc (see Sect. \ref{xray_sect}, and also \citealt{Turn18} and our Paper II by \citealt{Wang21}). Such internal reddening without neutral X-ray absorption was also seen in Seyfert-1 galaxies \object{MCG -6-30-15} \citep{Lee01}, \object{ESO~113-G010} \citep{Meh12}, and \object{IC~4329A} \citep{Meh18b}. Similarly, in these objects the internal reddening was explained by the lower-ionisation components of their warm-absorber outflows being dusty. In contrast to \citet{Cren01a}, who attribute the extinction and absorption to interstellar gas far from the nucleus in \ngc, we deduce that the internal reddening in \ngc can be attributed to the presence of dust in the warm-absorber outflows, likely being driven from the AGN torus. This dusty outflow scenario is further supported by the recent study of \citet{Alon19}, and references therein, which show in \ngc there are nuclear molecular outflows, as well as polar dust emission in the mid-IR, likely coinciding with the AGN ionisation cone.

We noted in Sect. \ref{uv_sect} that the HST/COS spectrum of \ngc contains an unreddened component in the far-UV (Fig. \ref{unred_fig}). An alternative explanation to this unreddened component may be a custom dust extinction law, one that is different from the ones considered in Fig. \ref{ext_fig}, including the SMC extinction with ${R_V = 4.0}$ that we used for de-reddening the spectra. Such a custom extinction curve would describe the intrinsic spectrum as a power law in $F_\lambda$ with index $\alpha=-7/3$ to match the red tail of the unreddened component. With such a custom extinction law the presence of the unreddened scattered continuum component is not essential. However, there are two points that favour the scattered radiation interpretation. Firstly, this component is needed to properly fit the optical/UV continuum when using a standard extinction law (Fig. \ref{unred_fig}). Secondly, the extended X-ray emission seen in the \chandra image of \ngc \citep{Zeng03}, and the extended NLR emission, imply the existence of a cone of scattered light similar to other nearby obscured Seyferts, such as NGC~4151 and \object{NGC~1068}. The study of \citet{Alon19} shows the presence of this cone in \ngc. Therefore, we deduce that the unreddened component detected in the HST COS spectrum is a real scattered emission component, likely originating from the extended NLR of \ngc.

\subsection{Broadband continuum emission of NGC 3227}
\label{cont_discuss}

The broadband continuum of \ngc is strongly modified by internal reddening and X-ray absorption. By modelling these effects in this paper, the underlying intrinsic emission of the AGN is uncovered. The NIR-optical-UV continuum of \ngc, measured thanks to broad HST spectral coverage, is fully consistent with a disk blackbody ({\tt dbb}) model. In order to fit the soft X-ray excess, an additional spectral component is required, which in this paper we modelled with a warm Comptonisation component ({\tt comt}). Our modelling of the \xmm and \nustar spectra shows that the hard X-ray spectrum of \ngc is consistent with a typical AGN X-ray power-law, produced in an optically-thin, hot, corona. The X-ray power-law is accompanied by a neutral, narrow, X-ray reflection component, producing the observed \FeKa line and the Compton hump at higher energies. 

We note that the broadband continuum model presented in this paper should be considered as only one feasible explanation. The primary purpose of Paper I is to establish the SED that is needed for the photoionization modelling of the outflows in \ngc, which are studied in more detail in our subsequent papers. To this end the single SED model presented in this paper is sufficient for the purpose of our investigation. Furthermore, considering the strong reddening in \ngc, and the caveats in modelling it, as well complexities in the strong X-ray absorption, \ngc is not the best candidate for evaluating merits of different models for the soft X-ray excess and thus tackling the long-standing issue in deciphering its origin in AGN.

From the {\tt dbb} model we can derive the inner disk radius and the accretion rate using the fitted parameters of the {\tt dbb} model. The radial temperature profile of the accretion disk is given by ${T(r) = \{ 3GM\dot M [ 1 - (R_{{\rm{in}}} /r)^{1/2} ] / (8\pi r^3 \sigma) \}^{1/4}}$ where $T$ is the local temperature at radius $r$ in the disk, $G$ the gravitational constant, $M$ the mass of the SMBH, $\dot M$ the accretion rate, $\sigma$ the Stefan-Boltzmann constant, and $R_{{\rm{in}}}$ is the radius at the inner edge of the disk. The flux at frequency $\nu$ from the disk seen by an observer at distance $d$ is given by ${F_\nu   = [4\pi h\nu ^3 \cos i/(c^2 d^2)] \int_{R_{{\rm{in}}} }^{R_{{\rm{out}}} } {({\rm{e}}^{h\nu /kT(r)}  - 1)^{ - 1}\, r\, {\rm{d}}r}}$ where $i$ is the inclination angle of the disk, $R_{{\rm{out}}}$ the radius at the outer edge of the disk, $h$ the Planck constant, and $c$ the speed of light. The parameter $R_{{\rm{out}}}$ cannot be constrained by spectral fitting since most of the radiation originates from the inner regions of the disk, thus $R_{{\rm{out}}}$ is fixed to $10^{3} R_{{\rm{in}}}$. The fitted parameters of the {\tt dbb} model are then the normalisation ${A = R_{{\rm{in}}}^2 \cos i}$ and the maximum temperature of the disk $T_{\rm{max}}$, which occurs at $r=(49/36)\times R_{{\rm{in}}}$. 

Therefore, using our best-fit normalisation $A$ value for the 2019 observation (Table \ref{continuum_table}), for which {\tt dbb} is better constrained thanks to HST spectral coverage, and the black hole mass of ${M = 5.96 \times 10^{6}}$~$M_{\odot}$ for \ngc \citep{Bent15}, we find ${R_{{\rm{in}}} = (11.3 / \sqrt{\cos i})\, R_{{\rm{g}}}}$, where ${R_{{\rm{g}}} = 2 G M / c^2}$. Adopting the inclination angle of $30^{\circ}$ from \citet{Alon19} for the nuclear ionisation cone of \ngc, we get ${R_{{\rm{in}}} = 12\, R_{{\rm{g}}}}$, with a propagated uncertainty of about $1\, R_{{\rm{g}}}$. This suggests the disk in \ngc is likely truncated. Using the above expression for $T(r)$, and our best-fit $T_{\rm max}$ (Table \ref{continuum_table}) and $R_{\rm in}$ value, the mass accretion rate $\dot M$ can be obtained, which is about 0.1~$M_{\odot}$ yr$^{-1}$ according to our {\tt dbb} model.

Our broadband continuum modelling of the 2019 observation gives an intrinsic bolometric luminosity of ${2.2 \times 10^{43}}$~\ergs. In the 2016 observation the intrinsic luminosity was higher by a factor of 2.1 than in the 2019 observation. The individual continuum components contribute to the bolometric luminosity in 2019 as following: 85\% from the disk blackbody ({\tt dbb}), 5\% from the warm Comptonisation component ({\tt comt}), 7\% from the X-ray power-law ({\tt pow}), and 3\% from the reflection component ({\tt refl}). Taking into account the black hole mass of 5.96~$\times 10^{6}$~$M_{\odot}$ \citep{Bent15}, the 2019 bolometric luminosity in \ngc corresponds to about 3\% of the Eddington luminosity. Our derived bolometric luminosity and the Eddington ratio matches the SED modelling results of \citet{Vasu10}: 2--3~${\times 10^{43}}$~\ergs and 2--4\% Eddington luminosity.

\section{Conclusions}
\label{conclusions}
In 2019 we found a transient obscuration event in Seyfert-1 galaxy \ngc using our \swift monitoring programme of a sample of AGN. Follow-up ToO observations were triggered with \xmm, \nustar, and HST. Here in Paper I we have determined the SED of \ngc, which is required for investigating the properties and origin of the new obscuring gas. We have modelled the broadband continuum of \ngc using \hst, \xmm, and \nustar spectra extending from NIR (10255 \AA) to hard X-rays (78 keV). We make use of both archival unobscured (2016) and new obscured (2019) spectra to disentangle the spectral components of the SED. We took into account all reddening and absorption processes that take place in our line of sight towards the nucleus of \ngc. Based on the results of our investigation we conclude the following points.
\begin{enumerate}
\item The underlying NIR-optical-UV continuum of \ngc is fitted well with a disk blackbody model with a maximum temperature of about 10~eV, inner disk radius of $12\, R_{{\rm{g}}}$, and an accretion rate of 0.1~$M_{\odot}$ yr$^{-1}$. The EUV continuum and the `soft X-ray excess' in \ngc are consistent with a warm Comptonisation component, up-scattering the disk seed photons in a warm, optically-thick corona.
\medskip
\item Our SED continuum modelling shows that the AGN radiation in \ngc is internally reddened with $\ebv = 0.45$. This reddening, which is seen in both the continuum and the Balmer lines, is most consistent with an SMC extinction law with $R_V=4.0$. We find the internal reddening in \ngc most likely occurs in dusty ionised outflows from the AGN torus, rather than being caused by neutral gas in the ISM of the host galaxy.
\medskip
\item Our modelling shows the presence of an unreddened far-UV continuum component at the shortest wavelengths of the COS spectrum. This component is likely scattered radiation from the extended NLR of \ngc. 
\medskip
\item From our modelling of the intrinsic broadband continuum in \ngc, we derive a bolometric luminosity of about ${2.2 \times 10^{43}}$~\ergs in 2019, which corresponds to about 3\% of the Eddington luminosity.
\end{enumerate}

\begin{acknowledgements}
This work is based on observations obtained with \xmm, an ESA science mission with instruments and contributions directly funded by ESA Member States and the USA (NASA). This research has made use of data obtained with the \nustar mission, a project led by the California Institute of Technology (Caltech), managed by the Jet Propulsion Laboratory (JPL) and funded by NASA. We thank the \swift team for monitoring our AGN sample, and the \xmm, \nustar, and \hst teams for scheduling our ToO triggered observations. This work was supported by NASA through a grant for \hst program number 15673 from the Space Telescope Science Institute, which is operated by the Association of Universities for Research in Astronomy, Incorporated, under NASA contract NAS5-26555. SRON is supported financially by NWO, the Netherlands Organization for Scientific Research. J.M. acknowledges the support from STFC (UK) through the
University of Strathclyde UK APAP network grant ST/R000743/1. S.B. acknowledges financial support from ASI under grants ASI-INAF I/037/12/0 and n. 2017-14-H.O. B.D.M. acknowledges support from Ramón y Cajal Fellowship RYC2018-025950-I. S.G.W. acknowledges the support of a PhD studentship awarded by the UK Science \& Technology Facilities Council (STFC). P.O.P. acknowledges financial support from the CNES French space agency and the PNHE high energy national program of CNRS. G.P. acknowledges funding from the European Research Council (ERC) under the European Union’s Horizon 2020 research and innovation programme (grant agreement No 865637). D.J.W. acknowledges support from STFC in the form of an Ernest Rutherford Fellowship. We thank the anonymous referee for the constructive comments.
\end{acknowledgements}

\end{document}